# Diamond-defect engineering of NV⁻ centers using ion beam irradiation

J. L. Sánchez Toural[1], J. García-Pérez[2], R. Bernardo-Gavito[2], D. Granados[2], A. Andrino-Gómez[1,3,4,5], G. García[4], J. L. Pau[1,5], M. A. Ramos[3,4,5,6,*],

and N. Gordillo[1,4,5,*]

[1]Laboratorio de Microelectrónica, Departamento de Física Aplicada, Universidad Autónoma de Madrid, E-28049 Madrid, Spain

[2]IMDEA Nanociencia, C/ Faraday, 9. E-28049, Madrid, Spain.

[3]Laboratorio de Bajas Temperaturas, Departamento de Física de la Materia Condensada, Universidad Autónoma de Madrid, E- 28049 Madrid, Spain

[4]Centro de Micro-Análisis de Materiales (CMAM), Universidad Autónoma de Madrid, C/ Faraday 3, E-28049 Madrid, Spain

[5]Instituto Nicolás Cabrera (INC), Universidad Autónoma de Madrid, E-28049 Madrid, Spain

[6]Condensed Matter Physics Center (IFIMAC), Universidad Autónoma de Madrid, E-28049 Madrid, Spain

*Corresponding authors

**Abstract**

The interplay between ion beam modification techniques in the MeV range and the controlled generation of negatively charged nitrogen-vacancy (NV⁻) centers in nitrogen-doped synthetic diamond crystals is explored. An experimental approach employing both light (H⁺) and heavy (Br⁺⁶) ions was followed to assess their respective impacts on the creation of NV⁻ centers, using different ion energies or fluences to generate varying amounts of vacancies. Photoluminescence spectroscopy was applied to characterize NV⁻ and neutral NV⁰ centers. Initially, no NV centers were detected post-irradiation, despite the presence of substitutional nitrogen and vacancies. However, after annealing at 800 ºC (and in some cases at 900 ºC), most samples exhibited a high density of NV⁰ and especially NV⁻ centers. This demonstrates that thermal treatment is essential for vacancy-nitrogen recombination and NV⁻ formation, often through electron capture from nearby nitrogen atoms. Notably, we achieved high NV⁻ densities without graphitization, which is essential for preserving the material's properties for quantum applications. This study underscores and quantifies the effectiveness of MeV-range ions in controlling vacancy distributions and highlights their potential for optimizing NV⁻ center formation to enhance the sensitivity of diamond-based quantum magnetic sensors.

---

* miguel.ramos@uam.es (Miguel Ángel Ramos)

## 1. Introduction

The physical properties of diamond crystals, such as color or electrical conductivity, can be controlled by doping impurities like B, N or P [1–13]. In particular, when diamond is doped with nitrogen, optically active nitrogen-vacancy (NV) centers can be induced under specific conditions. NV centers are point defects composed of a substitutional nitrogen adjacent to a vacancy in the diamond crystalline lattice. This center can be electrically neutral ($NV^0$ center) or can capture an additional electron, usually from a neighboring N atom, forming a negatively charged (NV⁻) center [14,15]. These NV⁻ centers have garnered significant attention due to their critical role in various quantum technologies. This type of defect is remarkable for its unique quantum properties, including sensitivity, temporal and spatial resolution, miniaturization tolerance, and operation across a wide range of temperatures. Additionally, its long spin coherence time, wide bandwidth and dynamic range, along with the ability to be optically initialized and read out [14,16,17], make it an ideal physical system for quantum sensing applications. [18,19]. These structures can be generated by strategically introducing defects into the diamond crystal lattice.

There are three primary irradiation techniques used to create nitrogen-vacancy centers in diamond: electron irradiation [10,20–24] , laser irradiation [25,26], and ion irradiation [13,24,27–30]. Each technique offers distinct advantages and disadvantages, making them suitable for specific applications depending on factors like desired NV⁻ center depth, density, and creation efficiency. In this work, we focused on ion irradiation due to its several advantages over laser and electron irradiation, such as precise control over concentration and depth, which make it suitable candidate to explore scalability to an industrial process. Following irradiation, thermal treatment is commonly employed to heal irradiation defects in the diamond lattice by facilitating the movement and recombination of substitutional atoms and vacancies [10,11,31]. This thermal treatment is crucial for enabling vacancy diffusion, promoting the formation of $NV^0$ complexes, and facilitating the charge state conversion to $NV^-$ centers [10,11,22,28]. Ultimately, this process results in a higher concentration of $NV^-$ centers.

Most of aforementioned approaches to produce $NV^-$ centers by ion irradiation employ either keV ion irradiation at diamond surfaces [28] or irradiation with GeV swift heavy ions in large facilities, producing columns of $NV^-$ centers along the deep ion track [27,30]. Recently, implantation of N ions at 2 MeV has also been employed [28], introducing nitrogen atoms and vacancies simultaneously. However, the required fluences needed are so high that graphitization is induced, which was overcome by irradiating on substrates above 550 ºC.

The aim of this work is to study alternative ways to controllably generate high density of $NV^-$ centers with MeV ions of quite different elements (H and Br), irradiating on commercial N-doped (~100–200 ppm) diamond crystals at room temperature and with fluences well below the graphitization threshold. After the work in Refs. [27,30], we know that GeV ion beams generate NV centers without annealing, whereas light ion and heavy ion beams at moderate energies need annealing. We have explored two very different ion beams with the aim of helping understand from a more fundamental point of view the mechanisms of NV center formation, besides seeking for the irradiation parameters that optimize the process. Our ultimate goal is that these defect-engineered $NV^-$ centers will be used to create a prototype quantum sensor for magnetoencephalography applications [19].

In Section 2, we present the type of samples employed and the experimental techniques used to carry out ion irradiation, thermal treatment and optical characterization. Three sets of experiments were conducted, which are consecutively described in Section 3, followed by our general conclusions in Section 4.

## 2. Materials and experimental methods
### 2.1. Samples

The samples used in this work for the irradiation purpose are type Ib synthetic diamond, one-side polished (3.0 mm × 3.0 mm × 0.5 mm) produced using the high-pressure, high-temperature (HPHT) synthesis process, acquired from Element Six™ [32]. According to the specifications, they contain less than 200 ppm nitrogen and are polished on one side with a {100} top face orientation.

Other four samples of synthetic diamond also supplied by Element Six[TM] (3.0 mm × 3.0 mm × 0.5 mm), were used as reference. Two of them (polycrystalline and monocrystalline) were manufactured using the HPHT synthesis process. As per specifications, they contain approximately 200 ppm nitrogen, exhibit green light absorption, and are brown in color, with an NV concentration of around 1 ppm. The other two reference samples (DNV[TM]1 and DNV[TM]14) were grown by chemical vapour deposition (CVD). They are single-crystal diamonds that contain deliberate and controlled levels of nitrogen-vacancy (NV) centers, specifically optimized for magnetometry applications, which are especially useful for system calibration and comparison with our samples. The reported concentration of NV centers is 300 ppb for DNV[TM]1 and 4.5 ppm for DNV[TM]14.

### 2.2. Optical characterization: Photoluminescence spectroscopy

The photoluminescence properties were studied at *IMDEA-nanociencia* using an optical cryostat attoDRY from Attocube GmbH [19], however all the measurements were performed at room temperature and pressure which are the requirements for the magnetometer. The confocal microscope is equipped with apochromatic objective, 0.65 mm working distance and 0.82 numerical aperture. The sample holder is mounted on an XYZ scanning stage. A scheme of the optical setup is displayed in Figure 1.

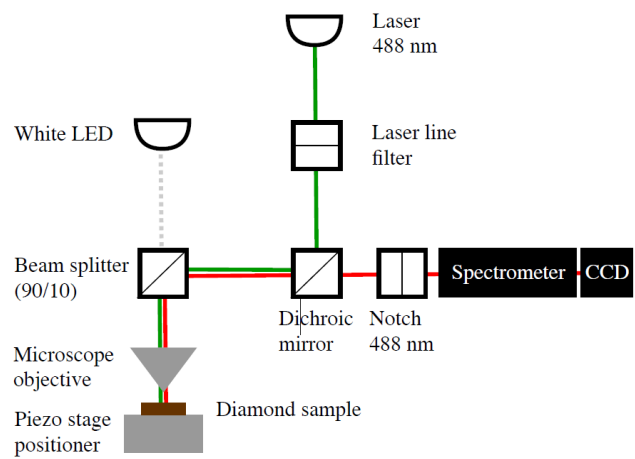

*Figure 1: Scheme of optical setup for the photoluminescence characterization.*

The optical setup for the photoluminescence characterization is based on a 488 nm argon-ion laser (Modu-Laser) with a maximum output power of 50 mW and a beam diameter of 0.65 mm (Figure 1). It includes a piezo-positioner, a microscope objective lens, a dichroic mirror (94:06), a 488 nm notch filter, a beam splitter (90:10), a CCD camera, a spectrometer and other optical elements as shown in the diagram. The laser power source is in the range between 40 μW and 2 mW which is partially absorbed by the dichroic mirror and the splitter, reducing the power incident on the sample to between 11.25 μW and 565 μW. The white LED is used for lighting and navigating the sample.

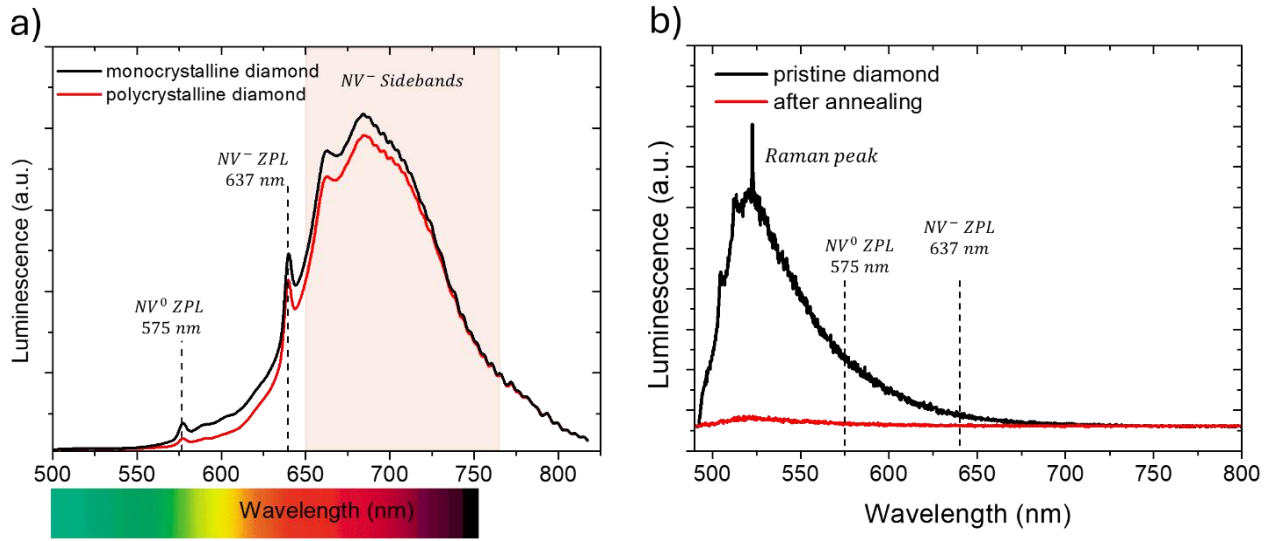

*Figure 2: a) Spectra smoothed out for an NV spin ensemble in a monocrystalline sample (black line) and polycrystalline sample (red line). b) PL spectra of Ib synthetic diamond before and after annealing.*

Figure 2a shows the characteristic photoluminescence spectra for an NV⁰ and NV⁻ ensemble embedded in both single-crystal and polycrystalline diamonds, which are remarkably similar. These curves illustrate the typical wavelengths involved, highlighting the zero-phonon lines (ZPL) and phonon sidebands. NV⁻ and NV⁰ centers can be distinguished by the different energy position of the zero phonon lines at 637 nm (1.94 eV) and 575 nm (2.15 eV), respectively [33]. The NV⁻ zero-phonon line is determined by the intrinsic difference in energy levels between the spin triplet ground and excited states. The NV⁻ phonon sideband shifts to a higher frequency during absorption, while in the case of fluorescence, it shifts to a lower frequency. This is to be compared with the case of pristine Ib diamond samples shown in Figure 2b, both before and after annealing, which are devoid of NV centers. The reference samples were characterized at room temperature under various conditions of laser light intensities with an integration time from 0.25 to 0.5 s, yielding nearly identical results. This enabled the subsequent characterization of irradiated samples using calibrated PL setups.

### 2.3. Ion irradiation and Monte-Carlo simulations

Diamond crystals were irradiated using both the standard and implantation (high vacuum) beam lines of the 5 MV electrostatic accelerator of tandem type in the Center for Micro Analysis of Materials (CMAM), located at the Universidad Autónoma de Madrid (UAM) [34].

To determine the theoretical density of vacancies, Monte-Carlo simulations were carried out with the SRIM-2013 software [35]. These simulations were run in the "detailed calculation with full damage cascades" mode with a displacement energy of 52 eV. The density of vacancies reflects the magnitude of lattice damage. Figures 3, 5 and 6 below show the depth profiles of vacancy density for the different employed fluences, ions, and energies. These profiles were calculated with a simple linear approximation without considering concurrent complex processes such as self-annealing, ballistic annealing and defect-defect interaction. We ignore any effect from electronic stopping, as was studied in detail and demonstrated in Ref. [36]. In that work, damage in diamond was studied for different irradiated ions and then with channeling, proving that structural damage correlated very well with nuclear stopping alone, without any effect from electronic stopping. Despite being an estimate, the calculated damage density can still be considered a reliable first-order approximation for evaluating the extent of lattice damage produced. This approach allows us to establish correlations between vacancy density produced by ion irradiation and subsequent $NV^-$ center creation. Consequently, three distinct experiments were proposed:

- Experiments I and II: We focused on exploring the implantation conditions for NV center creation. We investigated different factors such as the type of ion used, its implantation energy, and the total ion fluence.
- Experiment III: Refined analysis based on the results obtained from Experiments I and II. We specifically focused on improving the process for shallow NV center creation near the surface of the diamond using proton irradiation.

For Experiment I, we utilized a multipurpose standard chamber. This chamber offers flexibility for various experiments but may have limitations in defining the exact irradiation area. In contrast, Experiments II and III employed a dedicated implantation chamber. The implantation beam line offers superior control over the ion beam, allowing for more precise definition of the irradiated area and ensuring a more homogeneous distribution of implanted ions [34].

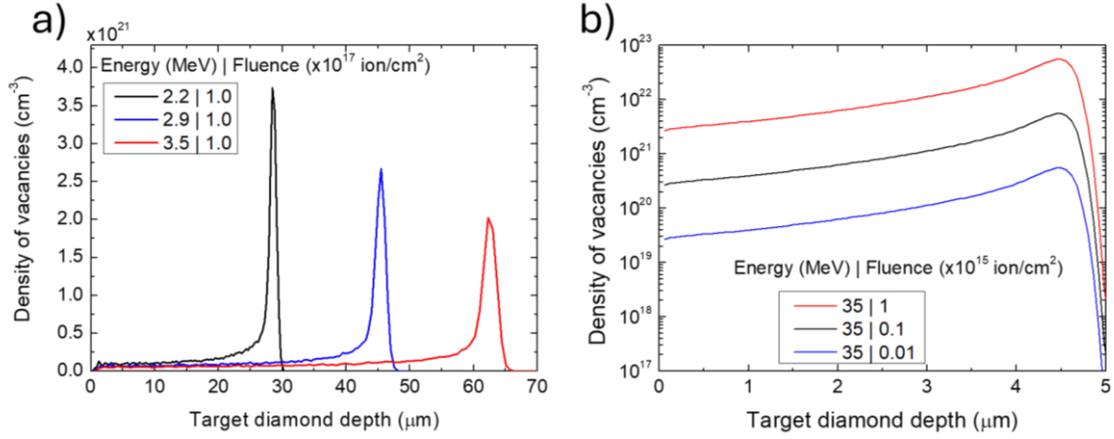

Figure 3: SRIM simulations for (a) H⁺ ion at different energies (2.2, 2.9 and 3.5 MeV) and for (b) Br$^{+6}$ at 35 MeV for three different fluences (10$^{15}$, 10$^{14}$ and 10$^{13}$ ion/cm$^2$).

In Experiment I, we aimed to differentiate between light and heavy ion irradiation. To achieve this, we irradiated three samples with light ions, such as H⁺, at identical fluences (×10$^{17}$ cm$^{-2}$) but at three different energies 2.2, 2.9 and 3.5 MeV. The objective of this experiment was to create vacancies at different depths (see Table 1 for details). Subsequently, we performed implantation on three additional samples using heavy ions, specifically Br$^{+6}$ at 35 MeV, but with varying fluences (1×10$^{13}$, 1×10$^{14}$ and 1×10$^{15}$ cm$^{-2}$). This part of the experiment aimed to generate different amounts of vacancies to investigate their influence on NV creation. The results of these simulations are represented in Figure 3 a) y b) respectively.

The density of vacancies estimated near the surface and near the Bragg peak for the experiment I are depicted in Table 1.

Table 1: Experiment I: Diamond samples implanted at different conditions for H⁺ and Br$^{+6}$ ions. The density of vacancies near the Bragg peak and near the surface, obtained from SRIM simulation, are indicated.

| Sample codes | Ion, energy | Fluence (ions·cm$^{-2}$) | Damage at Bragg peak (vac/cm$^3$) -- % [depth] | Damage at surface (vac/cm$^3$) -- ppm |
|---|---|---|---|---|
| H-I-1 | H⁺ 2.2 MeV | 1×10$^{17}$ | 3.8×10$^{21}$ – 2.2% @[28.5 μm] | 1.0×10$^{20}$ – 570 |
| H-I-2 | **H$^+$** 2.9 MeV | 1×10$^{17}$ | 2.7×10$^{21}$ – 1.5% @[45 μm] | 7×10$^{19}$ – 400 |
| H-I-3 | **H$^+$** 3.5 MeV | 1×10$^{17}$ | 2.1×10$^{21}$ – 1.2% @[62 μm] | 6×10$^{19}$ – 340 |
| Br-I-1 | **Br$^{+6}$** 35 MeV | 1×10$^{15}$ | 5.6×10$^{22}$ – 31.8% @[4.45 μm] | 3.5×10$^{21}$ – 20000 |
| Br-I-2 | **Br$^{+6}$** 35 MeV | 1×10$^{14}$ | 5.6×10$^{21}$ – 3.2% @[4.45 μm] | 3.5×10$^{20}$ – 2000 |
| Br-I-3 | **Br$^{+6}$** 35 MeV | 1×10$^{13}$ | 5.6×10$^{20}$ – 0.32% @[4.45 μm] | 3.5×10$^{19}$ – 200 |

In Experiment II, two samples were employed: one for H$^+$ and the other for Br$^{+6}$ ion irradiation. We lowered the irradiation energies and fluences for both protons (H) and bromine (Br) compared to Experiment I. This strategy targets shallower depths within the diamond lattice, placing the NV centers closer to the surface while also reducing the vacancy density. This adjustment is based on our observation that the number of generated NV centers was more than sufficient. Each sample was formally subdivided into four quadrants (Q$_X$, where $x$ ranges from 1 to 4) approximately 1.5×1.5 mm$^2$ each. These quadrants were subject to different conditions using an aluminium mask (0.5 mm in thickness) featuring a hole in one corner with a diameter of 1 mm. This approach enables us to expose different regions of the same sample to distinct irradiation conditions within a single experiment. In this case, we performed several irradiations using a multi-energy scheme. Specifically, for each sample quadrant, a series of different energies with correspondingly chosen fluences was employed, as described in Table 2. Figure 4 illustrates the subdivision of the sample into quadrants. Quadrants Q$_{1-3}$ were employed to adjust the energies and fluences with the objective of achieving a homogeneous pattern. Quadrant Q$_4$ remained non-implanted and served as a reference. The irradiation fluence increases progressively from Q$_1$ to Q$_3$, with each quadrant subjected to the irradiation energy profiles depicted in Figure 5, taking quadrants Q$_2$ as an example. In experiment II, we investigated a wider range of ion fluences for both H and Br irradiation. This allows us to optimize the number of vacancies created and potentially achieve a higher density of NV centers.

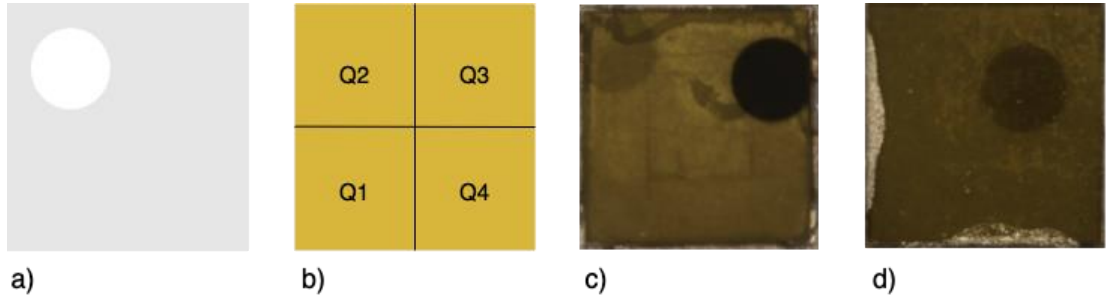

*Figure 4: a) Aluminium mask with a window to guarantee only a small circular section (1 mm diameter window) is irradiated. b) Quadrant distribution. c) Sample H-II after irradiation with $H^+$ shows how the regions implanted at fluences greater than $1 \times 10^{14}$ ions/cm$^2$ are clearly visible to the naked eye. d) Sample Br-II after irradiation with $Br^{+6}$ also shows dark areas caused by the irradiation process for the highest fluences.*

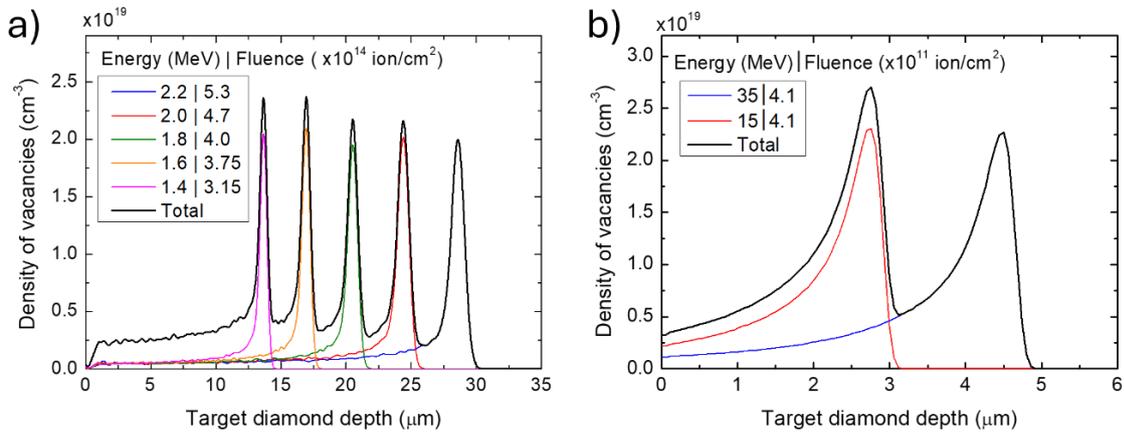

*Figure 5: SRIM simulations for (a) $H^+$ ion at different energies (from 1.4 to 2.2 MeV), corresponding to H-II-$Q_2$; (b) $Br^{+6}$ at 15 and 35 MeV, corresponding to Br-II-$Q_2$. Quadrants $Q_1$ and $Q_3$ present the same damage profiles, but multiplied by 1/10 and 10, respectively (see Table 2).*

The density of vacancies estimated near the surface and near the Bragg peak for Experiment II are depicted in Table 2.

*Table 2: Experiment II: Diamond samples implanted at different conditions for $H^+$ and $Br^{+6}$ ions. The density of vacancies near Bragg peak and near the surface for each quadrant, obtained from SRIM simulation, are indicated.*

| $H^+$ beam irradiation (sample codes: H-II-$Q_1$,$Q_2$,$Q_3$) | | | |
|---|---|---|---|
| Energy (MeV) | Fluence $Q_1$ (×10$^{13}$ cm$^{-2}$) | Damage at peaks $Q_1$ (×10$^{18}$ vac/cm$^3$) [depth] | Damage at surface $Q_1$ (×10$^{17}$ vac/cm$^3$) -- ppm $Q_2$ (×10$^{18}$ vac/cm$^3$) -- ppm |

| | $Q_2$ (×$10^{14}$ cm$^{-2}$) $Q_3$ (×$10^{15}$ cm$^{-2}$) | $Q_2$ (×$10^{19}$ vac/cm$^3$) [depth] $Q_3$ (×$10^{20}$ vac/cm$^3$) [depth] | $Q_3$ (×$10^{19}$ vac/cm$^3$) -- ppm |
|---|---|---|---|
| 2.2 | 5.30 | 2.0 @[28.5 μm] | |
| 2.0 | 4.70 | 2.2 @[24 μm] | |
| 1.8 | 4.00 | 2.2 @[20.5 μm] | 2.1 -- (1.2/12/120) |
| 1.6 | 3.75 | 2.3 @[17 μm] | |
| 1.4 | 3.15 | 2.3 @[13.5 μm] | |
| **Br$^{6+}$ beam irradiation** (sample codes: Br-II-$Q_1,Q_2,Q_3$) | | | |
| **Energy (MeV)** | **Fluence** $Q_1$ (×$10^{10}$ cm$^{-2}$) $Q_2$ (×$10^{11}$ cm$^{-2}$) $Q_3$ (×$10^{12}$ cm$^{-2}$) | **Damage at peaks** $Q_1$ (×$10^{18}$ vac/cm$^3$) [depth] $Q_2$ (×$10^{19}$ vac/cm$^3$) [depth] $Q_3$ (×$10^{20}$ vac/cm$^3$) [depth] | **Damage at surface** $Q_1$ (×$10^{17}$ vac/cm$^3$) -- ppm $Q_2$ (×$10^{18}$ vac/cm$^3$) -- ppm $Q_3$ (×$10^{19}$ vac/cm$^3$) -- ppm |
| 35 | 4.10 | 2.3 @[4.45 μm] | 3.5 -- (2/20/200) |
| 15 | 4.10 | 2.7 @[2.7 μm] | |

Finally, Experiment III: this stage explores the creation of NV centers near the surface using H$^+$ at 2.0 MeV irradiation. The experiment aimed to investigate the relationship between irradiation fluence and the resulting NV center density within the first few microns of the surface (see Figure 6). One of the samples was irradiated with a fluence of 2.23×10$^{15}$ cm$^{-2}$, while the other was with 1.12×10$^{16}$ cm$^{-2}$ fluence. This approach was adopted because, as will be discussed in detail in Section 3, the implantation with Br$^{+6}$ saturates the NV creation within the first few microns, making it challenging to achieve a uniform vacancy density near the surface. The density of vacancies estimated for each case near the surface for Experiment III is presented in Table 3, both being within the range of those produced in $Q_2$ and $Q_3$ of Experiment II.

Table 3: Experiment III: Diamond samples implanted at different conditions for H$^+$ ions. The density of vacancies near the surface, obtained from SRIM simulation are indicated.

| **H$^+$ beam irradiation – 2 MeV** |
|---|

| Sample codes | Fluence ($10^{15}$ cm$^{-2}$) | Damage at peak (vac/cm$^3$) | Damage at surface (vac/cm$^3$) -- ppm |
|---|---|---|---|
| H-III-1 | 2.23 | $1.2\times10^{20}$ | $2.2\times10^{18}$ -- 12.5 |
| H-III-2 | 11.2 | $6.0\times10^{20}$ | $1.1\times10^{19}$ -- 62.5 |

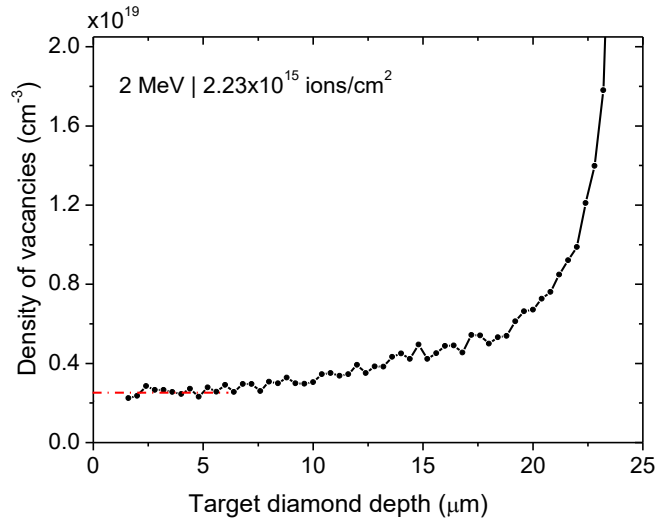

*Figure 6: SRIM simulations for H$^+$ at 2 MeV. The red line indicates the average density of vacancies produced few microns of the surface.*

### 2.4. Thermal treatments

Following irradiation, the samples were subjected to a thermal annealing treatment, aiming to activate the formation of NV$^-$ centers by promoting the recombination of vacancies with nitrogen atoms. Each step was characterized using photoluminescence (PL), as described above. These techniques allow us to assess the creation and quality of the NV centers.

Annealing treatment of the irradiated samples was carried out employing a quartz tube inside a commercial furnace. Previously, the diamond crystal was placed in a homemade vacuum-sealed quartz ampoule, which was carefully purged and then typically kept with 100–200 mbar of Ar gas at room temperature. Typically, the mobility of vacancies and C interstitial starts at 600°C, and the diffusion of vacancies becomes significant at 800°C [31]. Therefore, two temperatures (800 and 900 °C) were explored. Slow heating and cooling ramps were performed (more than 3.5 h each) with a dwell of 1 h for 800°C anneals in experiments I and II. Additionally, for experiment III, a second thermal treatment was conducted at 900°C, following the same outlined procedure.

## 3. Results and Discussion

For the reasons exposed above, we have performed different attempts and strategies to produce NV⁻ centers using MeV ions on N-doped commercial crystals of diamond. Earlier experiments using 1.1 GeV Au ions were able to detect the presence of NV⁻ centers without further annealing, in regions of high electronic stopping [27,30]. In contrast, we want to explore in which conditions one can use typical MeV ions (light protons of 2–3 MeV and heavier bromine ions up to 35 MeV) to create a high density of those defects, including the effect of post-annealing. We describe below the results of the three groups of experiments, whose irradiation parameters have been detailed in section 2.3.

### 3.1. Experiment I:

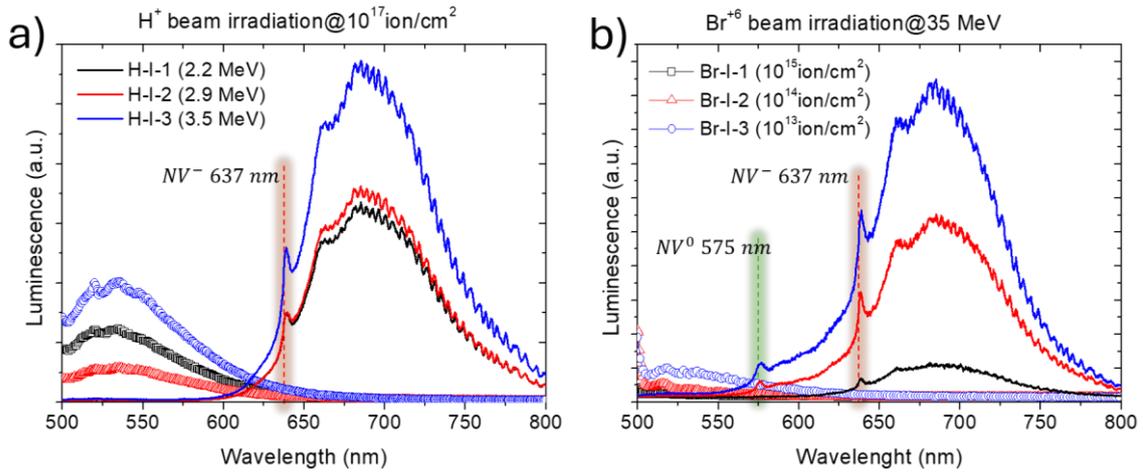

*Figure 7: PL characterization of samples a) H-I-1 (2.2 MeV-black), H-I-2 (2.9 MeV-red) and H-I-3 (3.5 MeV-blue) with proton irradiation; b) Br-I-1 ($10^{15}$ ion/cm²-black), Br-I-2 ($10^{14}$ ion/cm²-red) and Br-I-3 ($10^{13}$ ion/cm²-blue) with bromine irradiation. Open symbols represent the PL spectra of irradiated samples before thermal treatment while the solid lines represent the PL spectra after the thermal treatment of irradiated samples. The $NV^0$ and $NV^-$ lines are indicated.*

Figures 7a) and b) show the PL characterization of our diamond samples irradiated with protons and bromine, respectively. As one can see on the left side of the graphs (lower wavelengths), immediately after irradiation, there is no initial signature of the nitrogen-vacancy centers. When we subject these samples to a thermal treatment, the PL spectra on the right side in figures 7a) and b), represented with solid lines, reveal the emergence of new peaks. This demonstrates the crucial role of thermal annealing in two important processes: i) Recombination: the nitrogen atoms in the diamond lattice with the vacancies created by the irradiation; and ii) Electron Capture: some of these recombined NV centers capture an additional electron, transforming them into the negatively charged NV⁻ centers that we desire. Note that with Br we see both $NV^0$ and $NV^-$ centers, whereas with H, the $NV^0$ feature is absent or at least very weak.

We observe that for irradiation with both ions, the highest intensity of NV⁻ centers correspond to the experiment with the lowest vacancy density on the surface, H-I-3 and Br-I-3 (see Table 1), respectively, being on the order of $3–6×10^{19}$ vac/cm³.

### 3.2. Experiment II:

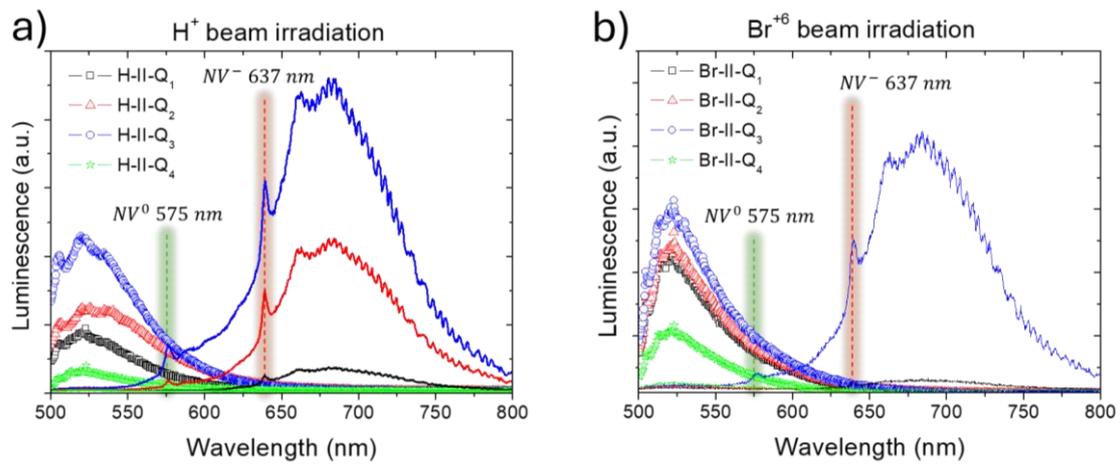

*Figure 8: PL characterization of samples a) H-II-Q₁ (black) ,H-II-Q₂ (red), H-II-Q₃ (blue) with proton irradiation and H-II-Q₄ (green) pristine diamond; b) Br-II-Q₁ (black) ,Br-II-Q₂ (red), Br-II-Q₃ (blue) with bromine irradiation and Br-II-Q₄ (green) pristine diamond Open symbols represent the PL spectra of irradiated samples before thermal treatment while the solid lines represent the PL measurements after the thermal treatment of irradiated samples. The $NV^0$ and $NV^-$ lines are indicated.*

Figures 8a) and b) show the PL characterization of our diamond samples irradiated with protons and bromine, respectively. One sample for each ion was used and each sample was divided into four quadrants to irradiate in multi-energy conditions but at different fluences, as described in Table 2. It should be noticed that the Q₄ was kept unirradiated as a reference, and was measured before and after annealing. As indicated above, immediately after irradiation there is no signature of nitrogen-vacancy centers. Only when we subject irradiated samples to annealing at 800°C, the PL spectra reveal the appearance of new peaks related to neutral and negative NV defects.

In this case, we are generating an equal or lower vacancy density near the surface compared to Experiment I. Our observations indicate that for both ions, the optimal results were obtained with the $Q_3$ samples (H-II-$Q_3$ and Br-II-$Q_3$ in Table 2), which had the highest fluence in Experiment II and were comparable to the optimal outcomes from Experiment I. In contrast, the $Q_1$ and $Q_2$ samples (H-II-$Q_1$,-$Q_2$ and Br-II-$Q_1$,-$Q_2$ in Table 2), which had fluences one to two orders of magnitude lower (and thus fewer surface defects), displayed reduced photoluminescence, due to fewer NV⁻ centers. In Experiment II, $NV^0$ centers are observed with both ions, although the concentration ratio [NV⁻]/[$NV^0$] is always high (see further discussion after Experiment III).

### 3.3. Experiment III:

This subsection explores the creation of NV centers near the surface only using 2 MeV proton irradiation. The experiment aimed to investigate the relationship between irradiation fluence and the resulting NV center density within the first 5 microns of the surface. For this particular experiment, we have studied 3 samples:

- H-III-0: as pristine sample (not irradiated but treated thermally) for comparison
- H-III-1: irradiated with protons at fluence of $2.25 \times 10^{15}$ cm⁻²
- H-III-2: irradiated with protons at five times the fluence of H-III-1

Two more reference samples from Element Six™ DNVB1 and DNVB14, with known low (300ppb) and high (4.5 ppm) NV concentrations, respectively, were also used in this part for comparison purposes.

The PL spectra of all studied samples in this section are shown in Figure 9. As observed, sample DNVB1 contains only $NV^0$ centers, while sample DNVB14 exhibits a higher number of NV⁻ centers compared to DNVB1. However, neutral NVº centers still predominate in DNVB14, and the sidebands of NV centers in the green region are more intense compared to our samples, which exhibit the expected behavior of diamond crystals with a few ppm of NV⁻ centers (see Fig. 2). It is important to note that these DNVB1 and DNVB14 samples were grown by CVD, unlike the other samples studied in this work, which were grown by HPHT. This suggests that CVD diamonds are more prone to generating neutral NVº centers rather than NV⁻ centers, in contrast to HPHT diamonds, at least in the case of reference samples fabricated by Element Six.

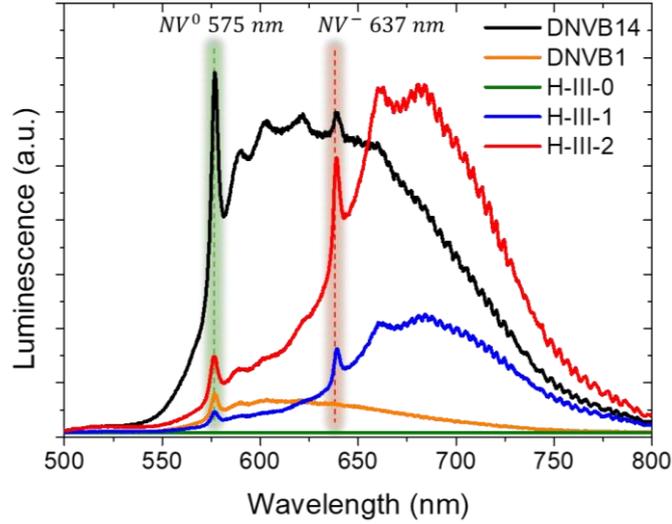

*Figure 9: PL spectra of the unirradiated sample (H-III-0) and irradiated samples (H-III-1 and H-III-2) after thermal annealing, compared to the reference samples DNVB1 and DNVB14.*

The concentration ratio $[NV^-]/[NV^0]$ has been estimated to be twice the ZPL intensity ratio, by using the difference in their Huang-Rhys factors for the two charge states [24]. Then, we can conclude that in the commercial samples DNVB1 and DNVB14 the concentration of $NV^0$ centers is higher than that of $NV^-$ centers, whereas in our irradiated diamond samples the concentration of $NV^-$ centers is one order of magnitude higher than that of $NV^0$ centers.

To assess the approximate density of $NV^-$ centers in absolute terms within the first microns of the irradiated samples (H-III-1 and H-III-2), we can make use of two previous works [22,24], where both the density of initial vacancies and that of $NV^-$ centers obtained after annealing at 800–850ºC irradiated diamond crystals with ~100–200 ppm of nitrogen, were measured using different approaches. Specifically, we refer to Fig, 5 of Ref. [22] and Table II of Ref. [24], where we will focus on data in the range of interest for us, namely those with initial vacancies about 10–100 ppm. Despite their different type of samples, irradiation conditions and experimental methods to determine the concentration of $NV^-$ centers, both works agree well. Using such "calibration curves", we estimate an $NV^-$ center density for sample H-III-1 of ≈1–2 ppm, reaching ≈10–15 ppm for sample H-III-2. Nevertheless, in both papers, *electron* irradiation was employed instead of ion irradiation, so this estimation should be taken with caution.

Finally, to study the influence of temperature during thermal treatment, we applied two annealing temperatures: 800°C and 900°C. The results, shown in Figure 10, indicate similar outcomes for both temperatures, suggesting that the increase in temperature within this range may not be critical for NV⁻ center formation, although the concentration ratio [NV⁻]/[NV⁰] appears to be influenced.

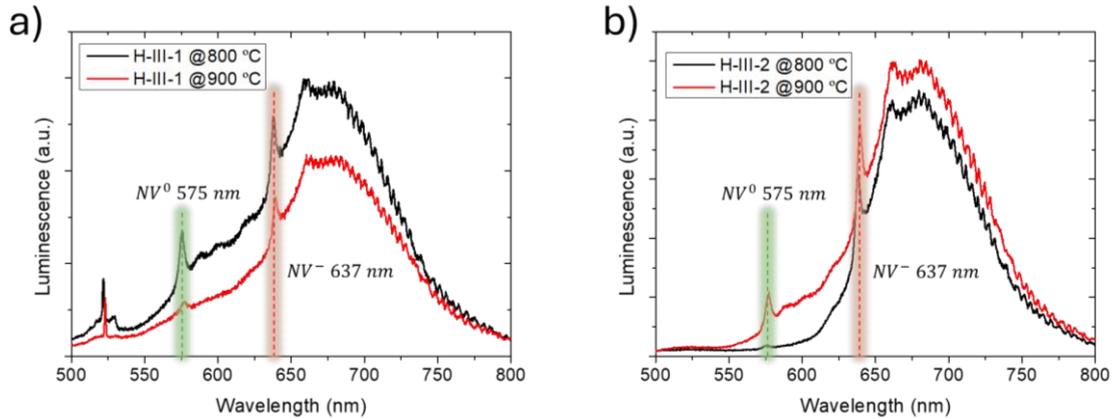

*Figure 10: PL spectra after thermal treatments at 800 and 900 °C for sample H-III-1 (a) and H-III-2 (b).*

All indications suggest that the optimal initial vacancy density is around $1\times10^{19}$ vac/cm$^3$, leading to the formation of approximately 10 ppm of NV⁻ centers, after annealing at 800–900ºC. Indeed, it makes sense that when approaching diamond crystals with ~100–200 ppm of substitutional nitrogen, the ideal number of generated vacancies to get a high density of NV⁻ centers should be lower than one half the number of N atoms (one needs at least one N atom for the NV center and another one nearby to provide the extra electron), but not too much lower. Therefore, a shallow density of vacancies ~$1\times10^{19}$ vac/cm$^3$ (~60 ppm of vacancies), which appears to generate ~10 ppm of NV⁻ centers using our experimental methods, seems to be the optimal parameters using MeV ions in Ib-type diamonds.

4. **Conclusion**

MeV ion irradiation proves to be an effective technique for creating NV⁻ centers in diamonds, offering precise control over depth and high-density generation. The process generates vacancies and interstitials, which can interact with pre-existing nitrogen impurities to form these centers. A comparison has been made between the NV⁻ center densities in irradiated and annealed samples versus reference samples. This comparison provides a clear measure of how effective the irradiation process is at generating NV⁻ centers and highlights areas for potential improvement. The annealing process plays a key role, as it facilitates the recombination of vacancies with nitrogen atoms in the diamond lattice, allowing for electron capture and the formation of NV⁻ centers.

For both light and heavy ion irradiations (at least for those studied in this work), the key parameter for NV⁻ center formation appears to be the vacancy density near the surface prior to thermal treatment. All indications suggest that the optimal initial vacancy density is on the order of $10^{19}$ vac/cm$^3$, leading to approximately ~10 ppm of NV⁻ centers. At higher densities, the damage becomes excessive, also in relation to the amount of available nitrogen, preventing efficient nitrogen-vacancy recombination. With lower densities, NV⁻ centers are detected, but the photoluminescence (PL) signal decreases, indicating reduced NV⁻ center formation. Our ion-irradiation method in type Ib synthetic diamond has also been found to provide a good conversion ratio NV$^0$ → NV⁻. We have shown that heavy ions such as Br can be used, though H+ seems to be more convenient given its more even distribution of vacancies in the most useful region a few μm behind the surface.

Thus, our findings not only highlight the versatility of ion beam techniques in the MeV range for diamond-defect engineering but also offer insights into the development of next-generation quantum technologies for advanced sensing applications.

**CRediT authorship contribution statement**

**J. L. Sánchez Toural:** Investigation, Formal analysis, Writing – review & editing. **J. García-Pérez:** Investigation, Formal analysis, Writing – review. **R. Bernardo-Gavito:** Investigation, Formal analysis, Writing – review. **D. Granados**: Conceptualization, Investigation, Visualization, Writing – review, Funding acquisition. **A. Andrino-Gómez:** Investigation, Formal analysis, Writing – review. **G. García:** Investigation, Writing – review & editing. **J. L. Pau:** Conceptualization, Investigation, Visualization, Writing – review & editing. **M. A. Ramos:** Conceptualization, Investigation, Visualization, Writing – original draft, Funding acquisition. **N. Gordillo:** Conceptualization, Investigation, Visualization, Writing – original draft, Funding acquisition.

**Declaration of competing interest**

We declare that we do not have any commercial or associative interest that represents a conflict of interest in connection with the work submitted.

**Acknowledgements**

We are grateful to the CMAM technical staff for their support and the beam time access with the proposal codes STD024/21, IMP005/22, IMP014/22, IMP015/23 and to Jose M. Castilla and Eduardo Ruiz for their help with experimental techniques. This research is supported by Spanish Ministry of Science, Innovation and Universities (MCIN/AEI/10.13039/501100011033) through projects PID2021-127498NB-I00 and PID2022-137779OB-C42, as well as by Severo Ochoa program for Centers of Excellence (CEX2020-001039-S) and by "María de Maeztu" program for Units of Excellence in R&D (CEX2023-001316-M). A.A.G. and J.G.P. thanks for the fellowships under contracts PRE2022-103066 and PRE2022-102081, respectively.

**Data availability**

All data are available in the main text, as well as from the corresponding author on reasonable request.